\documentclass{article}
\usepackage{geometry}                
\usepackage{graphicx}
\usepackage{stmaryrd}
\usepackage{amssymb}
\usepackage{amsthm, bm}
\usepackage{amsmath, cancel, centernot }
\usepackage[mathscr]{eucal}
\usepackage{mathtools}
\usepackage{epstopdf}
\title{Bell's Theorem, Non-Computability and Conformal Cyclic Cosmology: A Top-Down Approach to Quantum Gravity \\ \  \\ \large \emph{Written in celebration of Roger Penrose's 90th Birthday}}
\author{T.N.Palmer\\ Department of Physics, University of Oxford, UK\\
tim.palmer@physics.ox.ac.uk}
\date{\today}                                          
\makeatletter
\newcommand\be{\@ifstar{\[}{\begin{equation}}}
\newcommand\ee{\@ifstar{\]}{\end{equation}}}
\newcommand\bp{\begin{pmatrix}}
\newcommand\ep{\end{pmatrix}}

\newtheorem{theorem}{Theorem}

\makeatother
\begin{document}
\bibliographystyle{plain}
\maketitle
\begin{abstract}
This paper draws on a number of Roger Penrose's ideas - including the non-Hamiltonian phase-space flow of the Hawking Box, Conformal Cyclic Cosmology, non-computability and gravitationally induced quantum state reduction - in order to propose a radically unconventional approach to quantum gravity: Invariant Set Theory ({\rm IST}). In {\rm IST}, the fundamental laws of physics describe the geometry of the phase portrait of the universe as a whole: ``quantum'' process are associated with fine-scale fractal geometry, ``gravitational'' process with larger-scale heterogeneous geometry. With this, it becomes possible to explain the experimental violation of Bell Inequalities without having to abandon key ingredients of general relativity: determinism and local causality. Ensembles in {\rm IST} can be described by complex Hilbert states over a finite set $\mathbb C_p$ of complex numbers, where $p$ is a large finite integer. The quantum mechanics of finite-dimensional Hilbert spaces is emergent as a \emph{singular} limit when $p \rightarrow \infty$. A small modification to the field equations of general relativity is proposed to make it consistent with {\rm IST}.
\end{abstract}

\section{Introduction}
\label{intro}
Roger Penrose is one of the world's deepest thinkers. Like Einstein, he is convinced that quantum mechanics is only a provisional theory of physics and that we will need a paradigm shift to move forward. Nowhere is this more clearly stated than in his book 'The Large, the Small and the Human Mind' \cite{Penrose:1997} where he writes:
\begin{quote}
My own view is that to understand quantum non-locality we shall require a radically new theory. This new theory will not just be a slight modification to quantum mechanics but something as different from standard quantum mechanics as General Relativity is different from Newtonian Gravity. It would have to be something which has a completely different conceptual framework. In this picture, quantum non-locality would be built into the theory. 
\end{quote}

The purpose of this paper is to suggest a possible framework for such a ``radically new theory''. This framework draws on many of Penrose's ideas, notably the phase-space flow of Hawking Boxes, Conformal Cyclic Cosmology, non-computability, and the role of gravity in quantum state reduction. In this Introduction, we give a brief roadmap for this framework. 

Underlying everything is a conviction that the conventional interpretation of the experimental violation of Bell's inequality - that the laws of physics must either be indeterministic or violate local causality - lies at the heart of the seemingly insurmountable obstacles synthesising our theories of quantum and gravitational physics. Penrose has himself written about this on many occasions, noting that the violation of Bell's inequality seems at least in conflict with the spirit of relativity theory, even if the violation does not imply overt superluminal signalling. There is a widespread feeling that we haven't yet got to the bottom of this profound dilemma. 

In Section \ref {Bell} we propose an alternative interpretation of the experimental violation of Bell's inequality: violation of the Statistical Independence (SI) assumption. This assumption has been thoroughly misunderstood by the quantum foundations community. Sometimes it is claimed that violation of SI would mean that sub-ensembles of particle pairs in a Bell experiment are not statistically indistinguishable, leading to apocalyptic claims about the end of objective experimental science! Sometimes it is claimed that violation of SI would mean experimenters are not able to choose freely between measurement settings, that their brains have been subverted by the particles being measured! Both of these claims are incorrect. 

\begin{figure}
\centering
\includegraphics[scale=0.4]{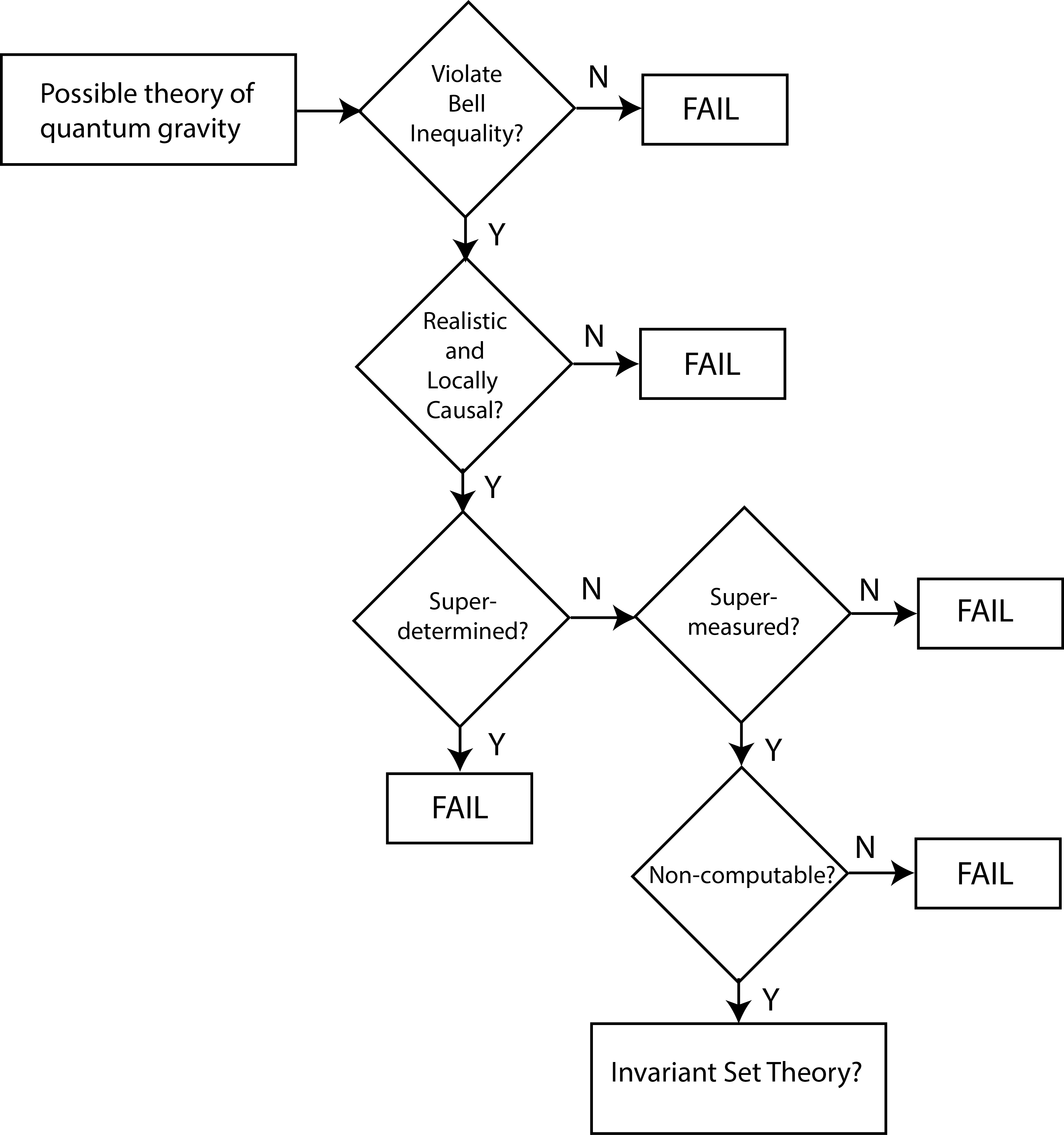}
\caption{\emph{A flowchart for some of the ideas discussed in this paper}}
\label{flowchart}
\end{figure}

In Section \ref{Bell} we note that SI can be violated (without any implications about the statistical equivalence of sub-ensembles or experimenter free choice) if there exists a non-trivial measure $\mu$ on  state space, such that on points where $\mu=0$, states are deemed inconsistent with the assumed laws of physics. An example where such a non-trivial measure occurs naturally is in dynamical systems where states are evolving on some dynamically invariant subset $I$ of state space.  If this subset is fractal, so that the dynamical system is chaotic, the subset has non-integer dimension and $\mu$ is Hausdorff. The fractal gaps in $I$ where $\mu=0$ correspond to specific counterfactual worlds which are inconsistent with the putative laws of physics. According to a theorem of Blum et al \cite{Blum}, many of the geometric properties of such fractal sets are algorithmically undecidable, i.e. non-computable. As we discuss, Penrose has often speculated that a quantum theory of gravity should be non-computable. In order to distinguish traditional ``superdetermined'' approaches to the violation of SI, here a violation of SI based on a non-trivial measure is referred to as ``supermeasured'' \cite{Hanceetal}. Fig \ref{flowchart} provides a flowchart for the logical progression of these ideas. In Fig \ref{flowchart}, the word ``realistic'' is assumed synonymous with deterministic. 

In Section \ref{InvariantSet} a theory of quantum physics is outlined based on the notion that the universe is evolving on such a cosmological invariant set $I_U$ - Invariant Set Theory ({\rm IST}) \cite{Palmer:2009a}, \cite{Palmer:2020}. In this theory, state-space trajectories have a fractal structure homeomorphic to the set of $p$-adic integers, for $p \gg 1$. Such a theory is very conceptually different from quantum mechanics - not least it implies that the quantum state vector has an ensemble intepretation. More dramatically, it implies that cosmology is not to be thought of as emergent from primitive laws on Planckian scales - the conventional approach where ``smaller'' is seen as inherently ``more fundamental''. Rather, the essential properties of quantum physics would arise from the geometry of the invariant set of the undivided universe.  This is what we mean in the title by a ``top-down'' \cite{Ellis} theory of quantum gravity. 

What evidence from cosmology is there for such a fractal invariant set $I_U$? In Section \ref{HB}, we discuss the Hawking Box, which Penrose discusses in a number of his books. Under plausible assumptions we show that the asymptotic end state of an isolated configuration of matter is a measure-zero fractal invariant set. Moving to a realistic model of the universe, we discuss whether Penrose's Conformal Cyclic Cosmology might similarly exhibit an asymptotic-in-time fractal invariant set.  

On the other hand, at least under non-gravitational circumstances, we  know that quantum theory is a supremely successful theory of physics. It is therefore important to understand the connection between {\rm IST} and quantum theory. To do this, in Section \ref{Complex}  we reframe quantum states as Hilbert vectors, not over the continuum $\mathbb C$, but over a set $\mathbb C_{p}$ of complex numbers $Ae^{i \phi}$, where $A^2$ and $\phi/2\pi$ are rationals of the form $m/p$ for integers $m$ and $p$, where $p>>1$. Such a granular Hilbert state space can be shown to provide a symbolic description of ensembles of trajectories in a $p$-adic neighbourhood of $I_U$, with possible links  back to earlier combinatoric ideas of Penrose's on spin networks. Just as $I_U$ is gappy and lacks the property of counterfactual definiteness, $\mathbb C_{p}$ is gappy and lacks the arithmetic closure properties of a full field.
 
A small modification to the field equations of general relativity is proposed to make it consistent with {\rm IST}. This modification may potentially help explain the phenomenon of dark matter.  In the Conclusions section, we discuss some future plans to develop {\rm IST}. 

Roger Penrose was internal examiner for my PhD in general relativity; my supervisor was Roger's close colleague and friend Dennis Sciama. On the occasions when I was lucky enough to take part in small discussion groups in Roger's office, I had a strong sense that the notion of cosmological homogeneity was being violated - that I was sitting in a very privileged frame of reference close to the centre of the universe. 

\section{Bell's Theorem and the Statistical Independence Assumption}
\label{Bell}

Our failure to synthesise quantum and gravitational physics may ultimately be due to the fact that we still haven't resolved the tension that exists between relativity theory and quantum theory, posed by the experimental violation of Bell inequalities. Specifically, general relativity is a deterministic locally causal theory, whilst, it is commonly asserted, deterministic locally causal theories of quantum physics must satisfy the (CHSH version of the) Bell Inequality
\be
\label{CHSH}
C(0,0)+C(0,1)+C(1,0)-C(1,1) \le 2
\ee
and thus be inconsistent with experiment. As usual, we imagine a source of entangled spin-1/2 particles prepared so that the total angular momentum of any pair of particles is zero. The spins of the particles are measured by remote experimenters Alice and Bob who can each choose to orient their measuring apparatuses in one of two ways (relative to a reference direction): conventionally these are referred to as $X=0$, $X=1$ (for Alice) and $Y=0$, $Y=1$ (for Bob). $C(X,Y)$ denotes the correlation in spin measurements for ensembles of particles, as a function of these detector settings. In the discussion below, $X'=1$ if $X=0$ and \emph{vice versa}, and similarly for $Y$.

However, in order to show that a locally causal deterministic theory satisfies (\ref{CHSH}), then it is also necessary to assume that theory satisfies the Statistical Independence assumption
\be
\label{SI}
\rho(\lambda |XY) = \rho (\lambda)
\ee
Here we presume that each pair of entangled particles is described by a supplementary variable $\lambda$, often referred to as a hidden variable - though in the uncomputable type of theory proposed here, there is no need for $\lambda$ to be hidden. For each $\lambda$ in a conventional hidden-variable theory, a value of spin (here $\pm 1$) is defined for each of the four values of $X$ and $Y$. $\rho(\lambda)$ is some probability density on the space of hidden variables. 

The assumption (\ref{SI}) is typically required to ensure that when the individual correlations in (\ref{CHSH}) are estimated from separate sub-ensembles of particle pairs (as necessarily happens in any real-world experimental test of the Bell inequality), then the hidden variables associated with these sub-ensembles are statistically equivalent to one another. In another interpretation, the assumption (\ref{SI}) is justified on the basis that experimenters' choices are not somehow subverted by the particles' hidden variables. A theory which violates (\ref{SI}) is referred to as superdeterministic \cite{HossenfelderPalmer}. As such, superdeterministic theories have typically been treated with contempt, and peremptorily dismissed. 

However, both of these interpretations of (\ref{SI}) are incorrect - they are not implied by a violation of (\ref{SI}). The reason the interpretations are incorrect relates to the definition of the  quantity $\rho(\lambda)$. As discussed in \cite{Hanceetal}, $\rho$ may depend both of factors which are under the control of the experimenter, and factors which are not. To make this explicit, we replace (\ref{SI}) by  
\be
\label{SI2}
\rho_{\rm Bell}(\lambda |XY) = \rho_{\rm Bell} (\lambda)
\ee
where $\rho_{\rm Bell}=\rho \times \mu$. We now treat $\rho$ as a probability distribution under the control of the experimenter: for example, it depends on the nature of the prepared state and whether the experiment was performed on Monday or Tuesday, and in London or New York. $\mu$, by contrast, is a non-trivial measure on state space, determined by the laws of physics. Points where $\mu=0$ describe states which are inconsistent with the assumed laws of physics. The actions and choices of experimenters, no much how freedom they have to do as they please, must nevertheless be consistent with the laws of physics. 

With this notation, the Statistical Independence assumption is actually (\ref{SI2}) not (\ref{SI}). Indeed, if ({\ref{SI}) is respected,  then it must be the case that
\be
\label{SI3}
\mu(\lambda |XY) = \mu(\lambda)
\ee
is violated. A theory where (\ref{SI}) is violated (with the new definition of $\rho$) will continue to be called ``superdetermined''. By contrast, a theory where (\ref{SI3}) is violated will be called ``supermeasured'' \cite{Hanceetal}. In the next Sections we describe a supermeasured theory of quantum physics.  

\section{Fractal Invariant Sets and Non-Computability}
\label{InvariantSet}

Consider the nonlinear differential equation
\be
\label{fp}
\dot r = r(1-r)
\ee
where $r > 0$. Starting from any initial $r \in \mathbb R^+$, it can easily be shown that $r$ will asymptotically approach $r=1$: $r=1$ is a fixed-point attractor of (\ref{fp}). We can add a second equation 
\be
\label{fp2}
\dot \phi =1
\ee
where $0 \le \phi < 2\pi$, in which case the asymptotic evolution of the dynamical system (\ref{fp}) and (\ref{fp2}), in the 2-dimensional state space described by the polar coordinate system $(r, \phi)$, will be the circle $r=1$. This is an example of a limit cycle. Such fixed points and limit cycles are examples of invariant sets: if a point lies on an invariant set, then its future evolution will always lie on it, and its past evolution will have always lain on it. Conversely, for a point which does not lie on  the invariant set, its future evolution will never will lie on it, and its past evolution has never has lain on it. The invariant measure $\mu$ associated with (\ref{fp}) is equal to 0 when $r \ne 1$. 

A more general class of invariant set is the fractal (or 'strange') attractor - an invariant set of a class of chaotic nonlinear systems. Perhaps the most famous of these is the Lorenz attractor (see Fig \ref{LorenzAttractor}) associated with the Lorenz equations \cite{Lorenz:1963}
\begin{align}
\frac{dX}{dt}&= \sigma (Y-X)    \nonumber \\
\frac{dY}{dt}&=X(\rho - Z) -Y    \nonumber \\
\frac{dZ}{dt}&=XY-\beta Z
\label{lorenzeqns}  
\end{align}

\begin{figure}
\centering
\includegraphics[scale=0.7]{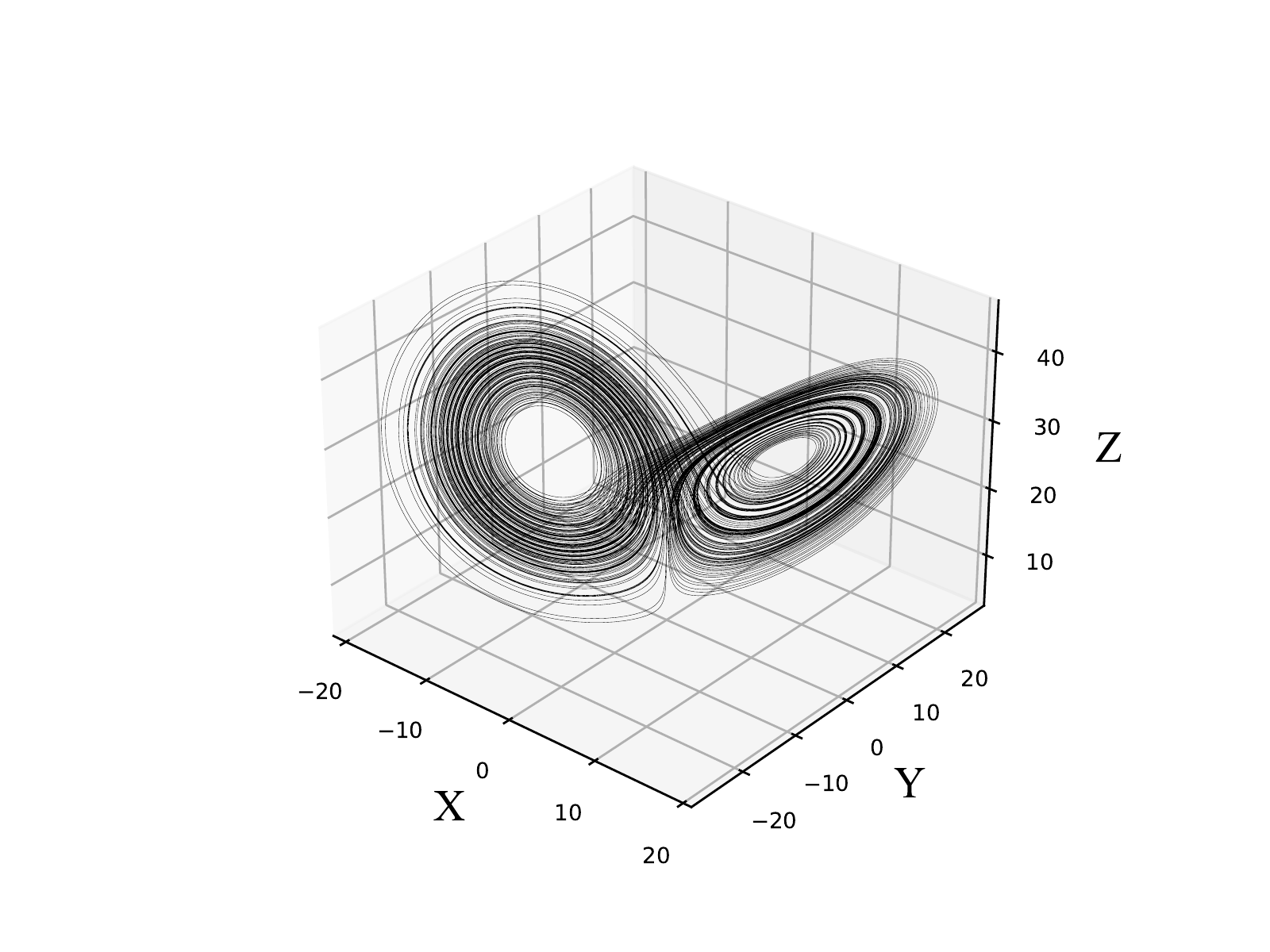}
\caption{\emph{The fractal invariant set of the Lorenz equations (\ref{lorenzeqns})}}
\label{LorenzAttractor}
\end{figure}

Such fractal attractors have a couple of important properties. Firstly they have zero volume in state space. That is to say, if an initial non-zero state-space volume is evolved in time using (\ref{lorenzeqns}), the volume will asymptotically shrink to zero. This means that there are processes which cause state-space velocity fields to be convergent. In the Lorenz equations, this process is described by time-irreversible dissipation in the model equations, and is associated with one or more negative Lyapunov exponents. As such, the invariant measure of the attractor is zero with respect to the Lebesgue measure of the full state space  in which the invariant set is embedded. 

To create a fractal structure, a simple state-space volume must be stretched and folded as it shrinks. In a chaotic system, the stretching is associated with instabilities with positive Lyapunov exponent. In the Lorenz system the negative Lyapunov exponent associated with dissipation dominates in magnitude over the positive Lyapunov exponent associated with instability. 

Fractal attractors provide another link to one of Penrose's key ideas. In \cite{Penrose:1997} Penrose discusses a toy universe whose states correspond to pairs of finite sets of polyominoes. Dynamical evolution in this toy universe is a non-computable rule, drawing on the result that there is no computational decision procedure for deciding when a polyomino set will tile the plane. Penrose writes: 
\begin{quote}
What about the actual universe? Well, I have argued [elsewhere] that there is something fundamental missing in our physics. Is there any reason from physics itself to think that there might be something non-computable in this missing physics? Well I think there is some such reason to believe this - that the true quantum gravity theory might be non-computable. 
\end{quote}

In his celebrated book The Emperor's New Mind \cite{Penrose:1989}, Penrose asks whether the fractal Mandelbrot Set is an example of a non-computable geometry. This possibility motivated the seminal book Complexity and Real Computation \cite{Blum} (co-authored by Steve Smale one of the pioneers of chaos theory). Blum et al set about answering the question of whether membership of the Mandelbrot Set is algorithmically decidable. Their argumentation applies equally to the fractal attractor $\mathcal A$ of a chaotic system.  We consider a putative algorithm/machine on the real numbers, which takes as input a point $\mathbf x$ in the state space of the chaotic system, and halts if $\mathbf x \in \mathcal A$. Blum et al's Path Decomposition Theorem implies that such an algorithm does not exist if $\mathcal A$ does not have integer dimension.  The very definition of a fractal is one whose Hausdorff dimension is not an integer. Hence we can conclude that indeed the fractal invariant sets of chaotic systems are uncomputable. This notion was further developed by Dube \cite{Dube:1993}, who showed that many of the classic undecidable problems of computing theory (e.g. the Post Correspondence  Problem) can be recast in terms of geometric properties of fractal attractors. 

We now come to the crucial point of this paper. Instead of defining a dynamical system by its differential equations (such as (\ref{fp}) or (\ref{lorenzeqns})), could we instead define it by the invariant set geometry, or, equivalently, by the invariant measure $\mu$? In the case of (\ref{fp}), this would be trivial. In the case of (\ref{lorenzeqns}) it is not trivial, precisely because the geometry is non-computational. However, there are techniques to characterise such fractal invariant sets topologically through what is called symbolic dynamics \cite{Gilmore}. Here one attaches symbolic labels, e.g. to the two lobes of the Lorenz attractor. The strings of symbolic labels associated with trajectory segments can be used to characterise the attractor topology. In this regard, it is worth recalling that Hilbert states in quantum mechanics are fundamentally symbolic in nature \cite{Schwinger} (we do not need to know precisely what is the difference between an alive and dead cat in order to define a superposed state of dead/alive cats). Here, differential equations (e.g. (\ref{lorenzeqns})) should merely be thought of as local representations of the global invariant set geometry. In particular, relative to this invariant set geometry, initial conditions and dynamical equations can no longer be considered independent of one another. 

In \cite{Palmer:2020} a model of quantum physics, referred to as Invariant Set Theory ({\rm IST}), was described, where state-space trajectories on an invariant set $I_U$ have a fractal structure isomorphic to the $p$-adic integers. Here $U$ denotes the (undivided \cite{BohmHiley}) universe, and it is assumed $U$ is evolving precisely on its fractal invariant set $I_U$. The fractal structure of trajectories on $I_U$ is illustrated in Fig \ref{IST}. 

\begin{figure}
\centering
\includegraphics[scale=0.3]{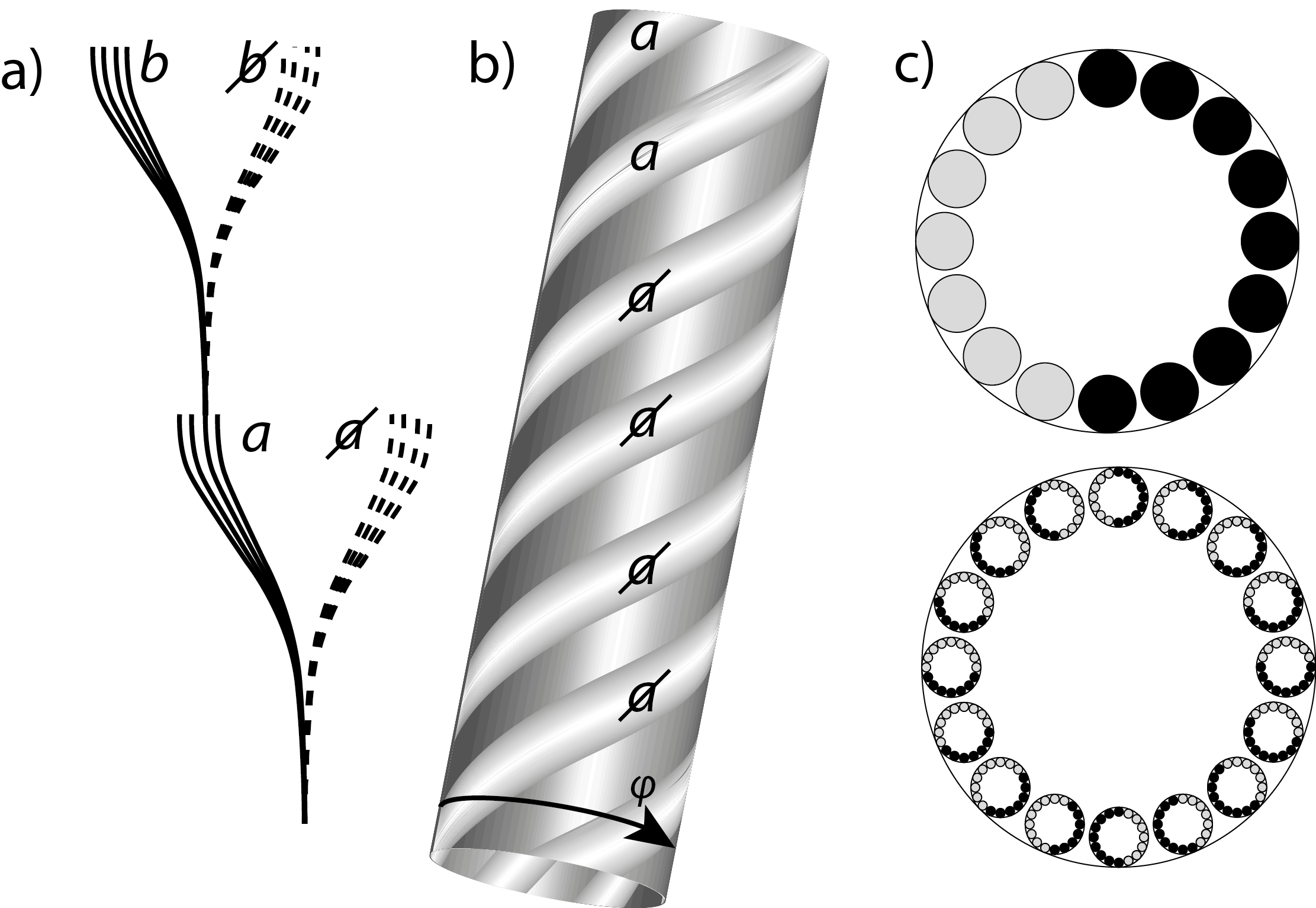}
\caption{\emph{A schematic of the local fractal state-space structure of the invariant set $I_U$ in Invariant Set Theory \cite{Palmer:2020}. a) An ensemble of trajectories decoheres into two distinct clusters labelled $a$ and $\cancel a$. Under a second phase of decoherence, this trajectory, itself comprising a further ensemble, decoheres into two further distinct clusters labelled $b$ and $\cancel b$. These clusters correspond to measurement eigenstates in quantum mechanics. b) Under magnification, a trajectory segment is found to comprise a helix of $p$ trajectories at the next fractal iterate. c) Top: a cross section through the helix of trajectories comprises $p$ (here $p=16$) disks coloured black or grey according to whether that trajectory evolves to the $a$ cluster or the $\cancel a$ cluster. Bottom: each of these $p$ disks itself comprises $p$ further disks coloured black or grey according to whether each trajectory evolves to the clusters $b$ or $\cancel b$. The fractal set of disks is homeomorphic to the set of $p$-adic integers. Reprinted with permission from Palmer T. N., Proc. R. Soc. A. 476, 20190350 (2020), under a Creative Commons 4.0 Attribution License.}}
\label{IST}
\end{figure}

Using the notion of Section \ref{Bell}, it can  be inferred from number-theoretic properties in {\rm IST}, (see \cite{Palmer:2020}) that, for example
\begin{align}
\mu (\lambda | X Y) \ne 0 \implies \mu (\lambda | X' Y) = \mu (\lambda | X Y')=0
\end{align}
That is to say, if the measurement settings for an actual Bell experiment on a particle pair with hidden variable $\lambda$ are $X$ and $Y$, then by definition the world characterised by the triple $(\lambda, X, Y)$ lies on $I_U$  and is associated with $\mu \ne 0$. In this situation, counterfactual worlds characterised by the triples $(\lambda, X', Y)$ and $(\lambda, X, Y')$ do not lie on $I_U$ and are associated with $\mu=0$. This means that it is possible to satisfy 
\begin{align}
\rho(\lambda | X Y) =\rho(\lambda | X' Y) = \rho(\lambda | X Y')=\rho(\lambda | X',Y')
\end{align}
while at the same time violating 
\begin{align}
\rho_{\rm Bell} (\lambda | X Y) =\rho_{\rm Bell}(\lambda | X' Y) = \rho_{\rm Bell}(\lambda | X Y')=\rho_{\rm Bell} (\lambda | X',Y')
\end{align}
Hence {\rm IST} violates Statistical Independence (\ref{SI2}) but not statistical independence (\ref{SI}). That is to say, the sub-ensembles of particle pairs which the experimenter chooses to perform his or her experiments are statically equivalent. Similarly, insofar as his or her choices do not fall foul of the laws of physics, experimenters are free to choose how to set their measuring apparatuses as they like: if they want to set their instruments based on the wavelength of light from distant quasars, so be it - there is nothing in {\rm IST} stopping them from so doing. Because {\rm IST} is non-computable, one cannot determine \emph{a priori} which putative states lie on $I_U$ and which not: this can only be asserted \emph{a posteriori}. Of course, it is one thing to say that {\rm IST} violates Bell's inequality, it is another thing to say it violates it as does quantum theory. We address this latter point in Section \ref{Complex}. 

Violations of Statistical Independence are sometimes criticised as being 'fine tuned'  \cite{WoodSpekkens}. However, by itself this criticism is meaningless: with respect to what measure or metric is the tuning fine? For example, the invariant measure $\mu(r)$ of the dynamical system defined by (\ref{fp}) is the delta function $\delta(r-1)$. From this point of view the precise value $r=1$ is far from being fine tuned, it is generic! Similarly, relative to the Hausdorff measure of $I_U$, points which lie on $I_U$ are similarly generic and not fine tuned. 

Indeed, as mentioned, $I_U$ is locally isomorphic to  the set of $p$-adic integers. Relative to the associated $p$-adic metric, a point which does not lie on $I_U$ is necessarily at least $p$-distant relative to a point which does lie on $I_U$ - even for pairs of points which, from a Euclidean perspective, appear extremely close. Hence, relative to measures and metrics which are naturally linked to the geometry of $I_U$, violations of (\ref{SI2}) are not fine tuned at all. And if the violation of (\ref{SI2}) is not fine tuned, it cannot be called conspiratorial, by definition. 

It is worth examining these issues from the perspective of the example introduced by Bell himself \cite{Bell}. Suppose the detector settings are determined by two pseudo-random number generators (PRNGs). In turn suppose the output of a PRNG is determined by the parity of the millionth digit of the input number. Consider a situation (a given $\lambda$) where the parity of both PRNG inputs was even. According to the discussion above, keeping $\lambda$ fixed, a putative world where one of the PRNG inputs was odd would not lie on $I_U$. 

This may violate one's intuition that the parity of the PRNG input is (to quote Bell \cite{BellFree}) ``unlikely to be the vital piece [of information] for any distinctively different purpose, i.e. it is otherwise rather useless''; that is to say, one's intuition may be that the rest of the universe couldn't care less about the parity of the PRNG inputs. Indeed, if these PRNG inputs weren't tied to a specific type of quantum experiment - a Bell experiment - then, perhaps the universe couldn't care less. However, when the PRNG inputs are linked to a quantum Bell experiment, then the universe does care and the counterfactual where the parity is perturbed, does not lie on $I_U$. Even Bell (at the end of \cite{BellFree}) acknowledges that his intuition in this matter may have been faulty. 

Is there any evidence that the universe may be evolving on a fractal invariant set in cosmological state space? We address this point in the next two sections. 

\section{Hawking Boxes and Conformal Cyclic Cosmology}
\label{HB}
We start with a description of a hypothetical system which Penrose calls the ``Hawking Box'' \cite{HawkingPenrose} \cite{Penrose:2004}, a vast (say galactic scale) box of matter whose walls are perfect mirrors, allowing no information to cross in or out. If we wait unimaginably long periods of time, then contents of the Box will vacillate irregularly between distributions of matter dominated by one or more black holes, and a field of almost pure Hawking radiation with no black hole

Penrose now considers the phase space/state space $\mathscr P_{\rm Hawking}$ of such a system.  This focus on state space is central to the ideas developed in the Sections above. A schematic of $\mathscr P_{\rm Hawking}$ is shown in Fig \ref{Hawking}. In the part of state space containing a black hole, Penrose assumes there will naturally be information loss, manifest as a convergence of the state-space velocity field (associated with a confluence of state-space trajectories). By contrast, in the part of radiation-dominated state space, quantum state reduction (in regions denoted ``R'') will cause state-space velocities to be divergent. 

It might seem strange at first sight to link the notion of state reduction to divergent flow. However, the construct merely illustrates the notion that, as normally envisaged, state reduction occurs when the quantum state evolves into one of a number of macroscopic possibilities. The divergent flow describes these different macroscopic possibilities. In the Everettian interpretation one imagines some kind of branching process where the quantum state literally splits into multiple macroscopically different worlds. Although Everettians like to think of this branching process as deterministic, the notion of a single state splitting into multiple states is the very epitome of indeterminism in nonlinear dynamical systems theory. In the framework developed in this paper, which is very much motivated by nonlinear dynamical systems theory, the divergent process illustrated in Fig \ref{Hawking} is assumed to represent an exponential divergence of trajectories like that in the left-hand panel of Fig \ref{IST}, where two trajectories never precisely lie on top of each other (contrasting with polynomial Everettian divergence). In Penrose's own writings, it is never quite clear which of these two paradigms he believes is most relevant in his collapse model. Here - consistent with strict determinism - we will assume that trajectory segments never intersect or lie on top of each other and that both the divergent and convergent flow describe exponentially increasing or decreasing distances between neighbouring trajectories. 

\begin{figure}
\centering
\includegraphics[scale=0.5]{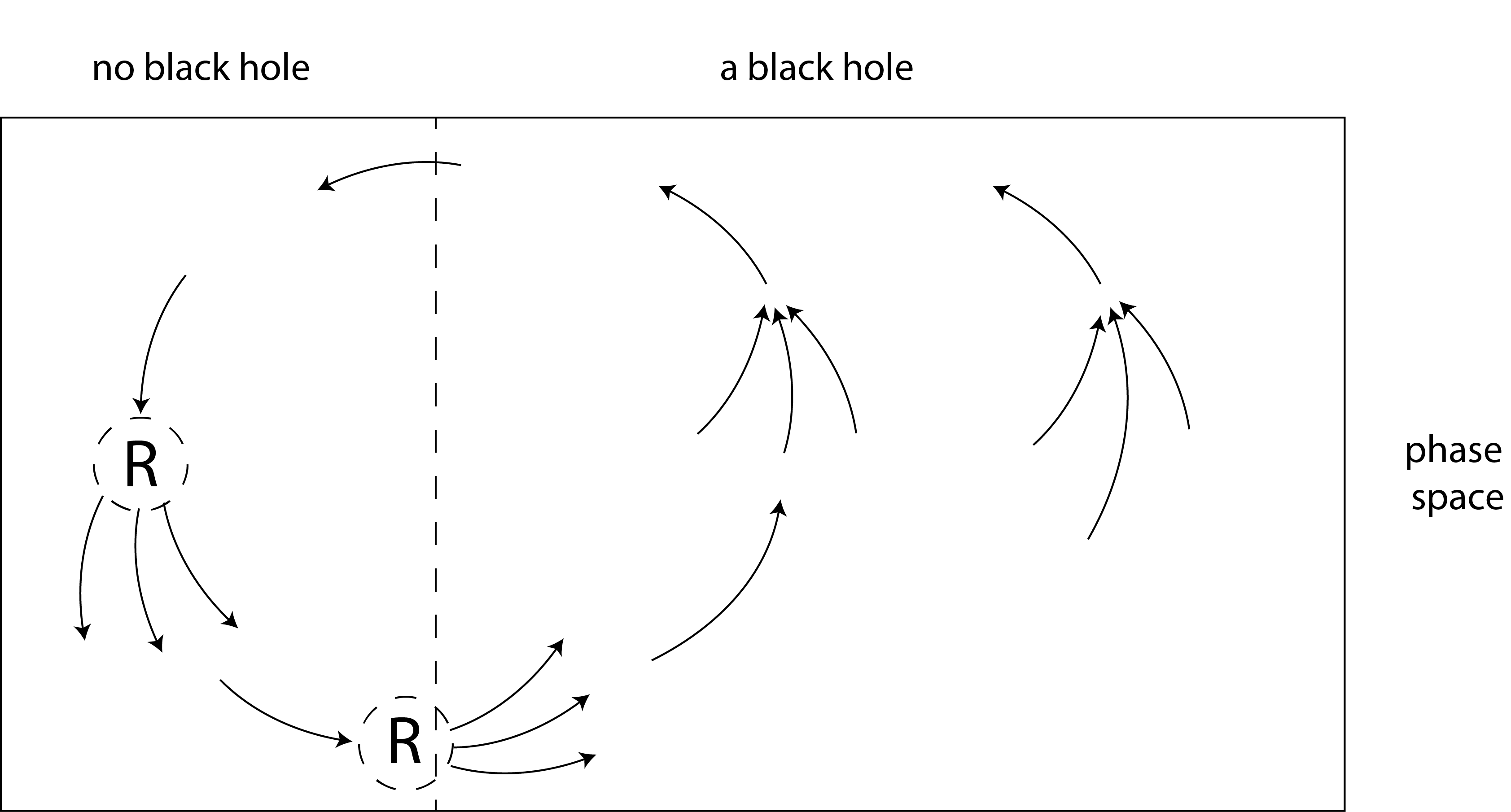}
\caption{\emph{The Hawking Box. Loss of phase-space volume occurs when a black hole is present. This may be (partially) balanced against regions where phase-space volumes grow due to quantum state reduction - here shown by the symbol ``R''. Adapted with permission from S. Hawking and R. Penrose, The Nature of Space and Time, Princeton Science Library, 2000. Copyright 2000 by Princeton University Press.}}
\label{Hawking}
\end{figure}

Because of these divergence and convergences, the phase-space flow is not Hamiltonian. It is strongly time asymmetric. Penrose postulates that the divergent and convergent processes are in balance so that ultimately there is phase-space volume preservation of the flow. However, here I we consider a possibility more consistent with chaos theory: that the convergence effect associated with black-hole dynamics (gravitational collapse more generally) wins out overall compared with the divergence effect associated with quantum state reduction. In this situation, the volume of the state-space flow will ultimately shrink to zero. This does not mean that the flow will shrink to a fixed point or limit cycle. The asymptotic invariant set of the Hawing box could be a zero volume (measure zero) fractal. 

Penrose discusses the Hawking Box in the context of the Liouville theorem. For a Hamiltonian system, the Liouville theorem demands that phase space volumes are conserved. In the asymptotic case where the phase-space volume of the Hawking Box is zero, this theorem will be obeyed. However, during the period of transient flow where volumes are shrinking, a generalised form of the Liouville theorem will be satisfied where probabilities, but not necessarily phase-space volumes, are conserved. This is an important point: loss of state-space volume does not necessarily imply a violation of conservation of probability. 

The convergence of the state-space flow during periods when the Hawking Box contains a black hole illustrates a feature of so-called final space-time singularities. Such final-time singularities are assumed to be dominated by Weyl curvature. The information in the degrees of freedom associated with the Weyl curvature is assumed lost when the black holes evaporate. Initial singularities, by contrast, can be expected to be free of Weyl curvature. As Penrose has explained in many of his books and papers, this asymmetry is needed to explain the Second Law of Thermodynamics.  The fundamental difference between initial and final singularities is encoded in Penrose's 'Weyl Curvature Hypothesis'. 

With colleague Paul Tod, Penrose realised that it was not actually necessary to make such a hypothesis. If the asymptotic end state of an epoch of the universe is pure zero rest-mass radiation whose origin was Hawking evaporation, then the universe will be conformally equivalent to an (initial) space-time singularity with finite (potentially zero) Weyl curvature. 

This led Penrose to abandon the Weyl Curvature Hypothesis, and instead postulate that our universe has a conformal cyclic structure \cite{Penrose:2010}. During CCC, the smooth matter fields at the beginning of a cosmological epoch clump gravitationally (and chaotically), ultimately forming black holes. These then evaporate leaving a pure radiation field (some decay of mass is postulated to ensure this final radiation field has zero rest mass and therefore satisfies conformal invariance). This final radiation field can be considered conformally equivalent to a new radiation dominated Big Bang. 

Let us consider the phase portrait of CCC. Like the Hawking Box, if gravitationally induced convergence wins out over divergence due to quantum state reduction, then we can assume that the asymptotic volume of the cosmological state-space invariant set of {\rm CCC} is zero. Again, like the Hawking Box, this does not mean we end up with a fixed-point invariant set. Instead, we will assume that this asymptotic end state is a zero-volume fractal invariant set in cosmological state space. 

The key idea here is that the existence of a non-computable fractal invariant set in cosmological state space {\rm IST}is consistent with multi-aeon cosmology, of which CCC is an example. 

\section{ Complex Numbers and Invariant Set Theory}
\label{Complex}

Clearly quantum theory is an extraordinarily successful theory. Yet here we are suggesting it be replaced by a deterministic ``supermeasured'' theory based the assumed fractal geometry of the invariant set $I_U$ of the universe. Just as general relativity theory is a fundamentally different theory to Newtonian gravity, the predictions of general relativity theory are the same as Newtonian gravity in the limit of large $c$. What is the equivalent ``weak limit'' of Invariant Set Theory, from which quantum theory would be emergent?

In quantum mechanics, states are represented by Hilbert vectors over the continuum field $\mathbb C$ of complex numbers, a mathematical object of great elegance, as Penrose has himself pointed out on many occasions (indeed Penrose's Twistor Theory is profoundly underpinned by the properties of complex holomorphic functions). However, is it possible the founding fathers of quantum mechanics were beguiled by such simplicity and elegance, ignoring the physical interpretational difficulties implied by complex Hilbert Space? I believe they were. The fact that $\mathbb C$ is a field means that all conceivable arithmetic operations on elements of $\mathbb C$ lead to elements which still lie in $\mathbb C$. However, a basic feature of a fractal set is that it is fundamentally gappy: arbitrary operations acting on a state which lies on an invariant set do not automatically lead to a state which lies on the invariant set, and indeed, typically they do not. 

Based on its fractal structure, {\rm IST} suggests an alternative to $\mathbb C$; specifically, it admits as physical only those wavefunctions $\psi = A e^{i \phi}$ where $A^2$ can be written in the form $m/p$ and $\phi$ is of the form $2\pi n/p$ where $m$, $n$ and $p$ are integers. Let $\mathbb C_{p}$ denote the set of such rational complex numbers. Since $\mathbb C_{p}$ is dense in $\mathbb C$ in the limit $p \rightarrow \infty$, it will be hard, for large $p$, to detect experimental differences between quantum mechanics and a theory based  on $\mathbb C_{p}$ directly. However, clearly $\mathbb C$ is a singular limit of $\mathbb C_p$: just as fractals are inherently ``gappy'', $\mathbb C_p$ is gappy relative to $\mathbb C$, no matter how large is $p$. 

As discussed in \cite{Palmer:2020}, elements of $\mathbb C_{p}$ have constructive representations as operators acting on bit strings of finite length $p$. To give a simple example, let $p=4$ and consider the bit string $S=\{a_1, a_2, a_3, a_4\}$ where $a_i \in \{1,-1\}$. Let $-S=\{-a_1, -a_2, -a_3, -a_4\}$. Define
\be
i^{1/2} S= \{-a_4, a_3, a_1, a_2\}
\ee
then it is easily shown that $i^2 S=-S$. In this way $i^{1/2}$ is a multiplicative 4th root of unity and can be written $e^{i \pi/4}$. $i$ itself is a square root of minus one. The formalism is readily generalised with bit strings of arbitrary length $p$, if $p$ is a power of 2. 

In this way, Hilbert vectors (and associated tensor products) over the set $\mathbb C_{p}$ can be interpreted \cite{Palmer:2020} as finite bit strings (and Cartesian products of bit strings). The bits themselves are interpreted as the symbolic labels of trajectories on $I_U$, where the labels correspond to the clusters to which trajectories evolve (see Fig \ref{IST}). These clusters correspond to measurement eigenstates of the relevant quantum observable. In this way, a Hilbert state over $\mathbb C_p$ represents an ensemble of trajectories (in a $p$-adic neighbourhood on $I_U$), where the ensemble is represented by the symbolic clustering labels. 

Because ensembles of trajectory segments in {\rm IST} can be described by complex Hilbert states over $\mathbb C_p$ then entanglement correlations in {\rm IST} will be, for all practical purposes, identical to those in quantum theory (except in the limit where the number of entangled qubits is comparable with $p$ whence entanglement correlations will become classical). Hence, for small numbers of entangled qubits,  {\rm IST} violates Bell's inequality exactly as does quantum theory.  

Consider the algebraic properties of $\mathbb C_{p}$. If $A_a e^{i \phi_a}$ and $A_b e^{i \phi_b}$ belong to $\mathbb C_{p/2}$, then $A_a A_b e^{i (\phi_a+\phi_b)}$ belongs to $\mathbb C_p$. But what about addition? From Euler's formula for complex numbers in $\mathbb C$ we have
\be
\label{add}
e^{i\phi_a} + e^{i\phi_b}= 2 \cos \frac{\phi_a- \phi_b}{2} e^{i \frac{\phi_a+\phi_b}{2}}
\ee
We now need to be mindful of Niven's Theorem \cite{Niven, Jahnel:2005}:
\begin{theorem}
\label{niven}
 Let $\phi/2\pi \in \mathbb{Q}$. Then $\cos \phi \notin \mathbb{Q}$ except when $\cos \phi =0, \pm \frac{1}{2}, \pm 1$. 
\end{theorem}
Hence if $\phi_a$ and $\phi_b$ are rational fractions of $2 \pi$, then $\cos (\phi_a - \phi_b)$ and hence $\cos^2 \frac{\phi_a- \phi_b}{2}$ are typically irrational. That is to say, the sum of $e^{i\phi_a} + e^{i\phi_b}$ does not belong to $\mathbb C_{p}$ even if the summands do. Unlike say the Gaussian  integers, $\mathbb C_{p}$ is not closed under both multiplication and addition. Because of this, $\mathbb C_{p}$ seems uninteresting from a mathematical point of view. However, from a physical point of view it is extremely interesting: strict superpositions of exponentials do not correspond to ensembles of trajectories on $I_U$. Moreover, (\ref{add}) implies
\be
\label{momentum}
-i \; \frac{e^{ik(x +\Delta x)}-e^{ik(x-\Delta x)}}{2\Delta x}= \frac{\sin (k\Delta x)}{\Delta x} \  e^{ikx}
\ee
for a spatial grid where $\ldots ,k(x- \Delta x), kx, k(x+ \Delta x), \ldots$ are all rational. Then, again by Niven's theorem, the square of the amplitude term on the  right hand side will typically not be rational, implying that subtraction of this type is undefined in $\mathbb C_{p}$. Conversely, if we choose $\Delta x$ so that the complex number on the right hand side of (\ref{momentum}) belongs to $\mathbb C_{p}$, then the individual complex exponents on the left hand side do not. 

However, in the limit $\Delta x \rightarrow 0$, subtraction of this type defines the momentum form of the quantum wavefunction. Basing our theory of quantum physics on $\mathbb C_{p}$ we have an immediate number-theoretic explanation of why the position and momentum of a particle are not simultaneously defined. One can put it like this. If the wavefunction describes an ensemble of trajectories where the position of a particle is measured, then a counterfactual ensemble of trajectories where the momentum of the same particle (same $\lambda$) is being measured, does not lie on the invariant set $I_U$. 

Is the loss of algebraic closure too heavy a price for such a simple physical interpretation? It might be if Hilbert vectors were fundamental. However, in a theory based on $\mathbb C_{p}$, these vectors clearly are not fundamental - they merely represent ensembles of trajectories on $I_U$.  The fundamental object here is the invariant set geometry itself. The Hilbert vectors merely describe statistical properties of this geometry. Algebraic closure arises at this deeper geometric level. For example, it is possible to add and multiply two $p$-adic integers and the result is a $p$-adic integer \cite{Katok}.

It would be interesting to reformulate Penrose's Twistor Theory in terms of $\mathbb C_p$ - twistor space, is, after all, a form of state space with complex structure. This might allow incorporation of Penrose's ideas about non-computability, state reduction and even CCC, into an approach to reformulating the laws of physics which has been the focus of Penrose's research since he was a young man.  

\section{Towards a Top-Down Theory of Quantum Gravity}
\label{QuantumGravity}

Conventional wisdom suggests that a theory of quantum gravity is needed primarily to explain the structure of space-time on the Planck scale. This fits naturally with the philosophy of Methodological Reductionism, where an attempt at providing an explanation at a deeper level inevitable involves a focus on smaller-scale entities. However, the discussion above suggests this philosophy may be profoundly misguided. 

Instead we propose some possible building blocks for a theory of quantum gravity, whose fundamental equations describe the phase-portrait geometry of the assumed fractal invariant set of a multi-aeon model of the universe, such as Conformal Cyclic Cosmology. Such a theory would  provide a clear delineation between quantum processes on the one hand, and gravitational processes on the other. Specifically, what we describe as ``quantum'' processes are those associated with the fractal structure of the invariant set, as illustrated in the middle and right-hand panels of Fig \ref{IST}. These are linked to positive local Lyapunov exponents on the invariant set (i.e. associated with locally diffluent trajectories) as an otherwise isolated system interacts with its environment. By contrast, what we describe as 'gravitational' processes are associated with larger-scale heterogeneities of the invariant set, such as the clustering process illustrated in the left-hand panel of Fig \ref{IST}. Such inhomogeneities link to negative Lyapunov exponents on $I_U$ (i.e. associated with confluent trajectories). Together, all the fundamental forces of nature, including gravity, will be describable in terms of the total (stable and unstable) geometry of $I_U$. As yet, no clear ``picture'' of this global geometry has emerged. 

This delineation between ``quantum'' and ``gravity'' suggests there is no reason to think of gravity as some force subject to the rules of quantum field theory. In {\rm IST} there is no such thing as a graviton. Moreover, there is no reason for gravity to be a 'witness of quantum entanglement' \cite{Bose} \cite{Vedral}.

In {\rm IST}, the convergence of trajectories into a finite number of clusters is, by our definition, an inherently gravitational process. This, therefore, is consistent with Penrose's notion of gravitationally induced collapse of the wavefunction (``R'' in Fig \ref{Hawking}). The key difference with Penrose's mechanism is that in {\rm IST}, there is no continuous 'wavefunction' from which collapse takes place. Instead, by describing quantum states as vectors over $\mathbb C_{p}$ rather than $\mathbb C$, then Hilbert states simply describe finite ensembles on $I_U$.  In this sense, the measurement problem is rather trivially solved in {\rm IST}. 

{\rm IST} suggests that one should generalise the equations of general relativity to 
\be
\label{GRmod}
G_{\mu \nu} (\mathcal M)= \frac{8 \pi G}{c^4} \int_{\mathcal M'}  T_{\mu \nu}(\mathcal M') \; \mu_p(\mathcal M', \mathcal M)\; d \mu
\ee
where $\mathcal M$ denotes space-time corresponding to our trajectory on $I_U$, and $\mathcal M'$ denotes space-times corresponding to other trajectories on $I_U$.  $\mu_p$ is a measure on $I_U$ which 'smears out'  the singular delta function $\delta(\mathcal M' - \mathcal M)$ of general relativity onto a $p$-adic neighbourhood of $I_U$ - Einstein always believed the right hand side of his field equations was somewhat provisional. That is to say, consistent with the primacy of the geometry of $I_U$, the curvature of space-time feels the energy-momentum of matter fields in space-times associated with neighbouring trajectories on $I_U$, in addition to the energy-momentum of matter fields on the trajectory corresponding to our space-time. This extra source of space-time curvature would manifest itself as if there were some form of ``dark matter'' in our space-time (which, according to {\rm IST}, there isn't). Additionally, compared with $\delta(\mathcal M' - \mathcal M)$, the non-singular nature of $\mu_p(\mathcal M', \mathcal M)$ would likely prevent the blow-up of space-time curvature at space-time singularities in $\mathcal M$ - a requirement for any putative theory of quantum gravity. 

Suppose indeed the quantum-matter fields on $I_U$ are associated with the (positive Lyapunov subspace of) the geometry of $I_U$. Then (\ref{GRmod}) describes a relationship between the pseudo-Riemannian geometry of space-time and the $p$-adic geometry of $I_U$. If this programme of research is successful, we will have reduced \emph{all} of physics to geometry. 

An experimental test of {\rm IST} \cite{Hance} relates to the fact that the bit strings associated with complex Hilbert states have finite length $p$. This implies that only a finite number of qubits can be entangled and still show quantum entanglement statistics. This inherent limit of quantum coherence might be intimately related to Penrose's own ideas on gravitational decoherence, given the discussion about gravity and the heterogeneity of $I_U$.

\section{Conclusions}
\label{Conclusions}

Roger Penrose, more than anyone, has taught us not to be afraid to think ``out of the box''. This has served me well in my professional career. Here I have returned to the topic I studied for my doctoral thesis under the guidance of both Roger Penrose and Dennis Sciama, and have suggested some ``out of the box'' ideas for synthesising quantum and gravitational physics. These ideas are motivated by ideas that Penrose himself has promoted: notably cosmic cyclic cosmology, non-computability and gravitationally induced quantum state reduction. This has led to a proposal for a future quantum theory of gravity which completely deviates from the usual philosophy of methodological reductionism, where ``smaller is more fundamental''. The key is the phase-portrait geometry of the invariant set $I_U$ of the undivided \cite{BohmHiley} universe. Relative to such a geometry, the basic concepts of determinism and causality which underpin general relativity, still hold, and yet Bell's inequalities can be violated as they are in quantum theory. Although we have formulated a small modification to the field equations of general relativity, by far the larger change will be to the equations of quantum physics, where Hilbert states are no longer seen as fundamental, but are merely statistical descriptions of strings of symbolic labels given to state-space trajectories. The new framework for quantum physics is indeed ''as different from standard quantum mechanics as general relativity is from Newtonian gravity'' as Penrose wrote that it would have to be. In particular, quantum mechanics emerges as a \emph{singular} limit of Invariant Set Theory {\rm IST}. 

Of course, the ideas proposed here need considerable further development. So far,  {\rm IST} has only been developed to describe finite-dimensional Hilbert States. A crucial question concerns the invariant set interpretations of continuous quantum wavefunctions $\psi(x,t)$ in physical space-time. This is needed to explore the concept of wave-particle duality, not discussed in this paper, from an invariant set perspective. The next paper on {\rm IST} will develop this topic. A second paper will explore the extent to which the revision (\ref{GRmod}) of the Einstein field equations can indeed explain the phenomenon of dark matter.  

It is an honour to write this paper to celebrate Roger's 90th Birthday. 

\section*{Data Availability Statement}

Data sharing is not applicable to this article as no new data were created or analysed in this study. 

\bibliography{mybibliography}

\begin{thebibliography}{10}

\bibitem{BellFree}
J.S. Bell.
\newblock {\em Free variables and local causality. In `Speakable and
  unspeakable in quantum mechanics'}.
\newblock Cambridge University Press, 1993.

\bibitem{Bell}
J.S. Bell.
\newblock {\em Speakable and unspeakable in quantum mechanics}.
\newblock Cambridge University Press, 1993.

\bibitem{Blum}
L.~Blum, F.Cucker, M.Shub, and S.Smale.
\newblock {\em Complexity and Real Computation}.
\newblock Springer, 1997.

\bibitem{BohmHiley}
D.~Bohm and B.J.Hiley.
\newblock {\em The Undivided Universe}.
\newblock Routledge, 2003.

\bibitem{Bose}
S.~Bose, A.Mazumdar, G.W.Morley, H.Ulbricht, M.Toros, M.~Paternostro, A.Geraci,
  P.Barker, M.Kim, and G.Milburn.
\newblock A spin entanglement witness for quantum gravity.
\newblock {\em Phys.Rev.Lett}, 119:240401, 2017.

\bibitem{Dube:1993}
S.~Dube.
\newblock Undecidable problems in fractal geometry.
\newblock {\em Complex Systems}, 7:423--444, 1993.

\bibitem{Ellis}
G.F.R. Ellis.
\newblock Top-down causation and quantum physics.
\newblock {\em Proceedings of the National Academy of Sciences},
  115:11661=11663, 2018.

\bibitem{Gilmore}
R.~Gilmore and M.~Lefranc.
\newblock {\em The Topology of Chaos}.
\newblock Wiley, 2002.

\bibitem{Hance}
J.~R. Hance, T.N.Palmer, and J.Rarity.
\newblock Experimental tests of invariant set theory.
\newblock arXiv:2102.07795, 2021.

\bibitem{Hanceetal}
J.~R. Hance, T.N.Palmer, and S.Hossenfelder.
\newblock Supermeasured: how to violate {S}tatistical {I}ndependence without
  violating statistical independence.
\newblock arXiv:2108.07292, 2021.

\bibitem{HawkingPenrose}
S.~Hawking and R.~Penrose.
\newblock {\em The Nature of Space and Time}.
\newblock Princeton Science Library, 2000.

\bibitem{HossenfelderPalmer}
S.~Hossenfelder and T.N. Palmer.
\newblock Rethinking superdeterminism.
\newblock Frontiers in Physics, to appear. arXiv:1912.06462, 2019.

\bibitem{Jahnel:2005}
J.~Jahnel.
\newblock When does the (co)-sine of a rational angle give a rational number?
\newblock arXiv:1006.2938, 2010.

\bibitem{Katok}
S.~Katok.
\newblock {\em p-adic Analysis compared with Real}.
\newblock American Mathematical Society, 2007.

\bibitem{Lorenz:1963}
E.N. Lorenz.
\newblock Deterministic nonperiodic flow.
\newblock {\em J.Atmos.Sci.}, 20:130--141, 1963.

\bibitem{Vedral}
C.~Marletto and V.~Vedral.
\newblock An entanglement-based test of quantum gravity using two massive
  particles.
\newblock {\em Phys. Rev. Lett.}, 119:240402, 2017.

\bibitem{Niven}
I.~Niven.
\newblock {\em Irrational Numbers}.
\newblock The Mathematical Association of America, 1956.

\bibitem{Palmer:2009a}
T.N. Palmer.
\newblock The invariant set postulate: a new geometric framework for the
  foundations of quantum theory and the role played by gravity.
\newblock {\em Proc. Roy. Soc.}, A465:3165--3185, 2009.

\bibitem{Palmer:2020}
T.N. Palmer.
\newblock Discretization of the {B}loch sphere, fractal invariant sets and
  {B}ell's theorem.
\newblock {\em Proc. Roy. Soc.}, https://doi.org/10.1098/rspa.2019.0350,
  arXiv:1804.01734, 2020.

\bibitem{Penrose:1989}
R.~Penrose.
\newblock {\em The Emperor's New Mind: Concerning Computers, Minds, and the
  Laws of Physics}.
\newblock Oxford University Press, 1989.

\bibitem{Penrose:1997}
R.~Penrose.
\newblock {\em The Large, the Small and the Human Mind}.
\newblock Cambridge University Press, 1997.

\bibitem{Penrose:2004}
R.~Penrose.
\newblock {\em The Road to Reality: A Complete Guide to the Laws of the
  Universe}.
\newblock Jonathan Cape, London, 2004.

\bibitem{Penrose:2010}
R.~Penrose.
\newblock {\em Cycles of Time: An Extraordinary New View of the Universe}.
\newblock The Bodley Head, 2010.

\bibitem{Schwinger}
J.~Schwinger.
\newblock {\em Quantum Mechanics: Symbolism of Atomic Measurements}.
\newblock Springer, 2001.

\bibitem{WoodSpekkens}
C.J. Wood and R.W.Spekkens.
\newblock The lesson of causal discovery algorithms for quantum correlations:
  Causal explanations of bell-inequality violations require fine tuning.
\newblock {\em New J. Phys.}, 17:033002, 2015.

\end{thebibliography}

\end{document}